\documentclass[11pt]{article}

\usepackage[preprint]{acl}

\usepackage{times}
\usepackage{latexsym}

\usepackage[T1]{fontenc}

\usepackage[utf8]{inputenc}

\usepackage{microtype}

\usepackage{hyperref}
\usepackage{url}
\usepackage{booktabs}
\usepackage{tablefootnote}
\usepackage{array}
\usepackage{graphicx}
\usepackage{pgfplots}
\usepackage{float}
\pgfplotsset{compat=1.18}
\usepgfplotslibrary{groupplots}
\usetikzlibrary{calc,arrows.meta}
\definecolor{commitblue}{HTML}{1F5AA6}
\definecolor{selfpurple}{HTML}{7A5195}
\definecolor{openorange}{HTML}{E17C05}
\definecolor{zeroteal}{HTML}{2A9D8F}
\definecolor{debategreen}{HTML}{4C956C}
\definecolor{indgray}{HTML}{7A7A7A}
\definecolor{syntheticbg}{HTML}{FFF8E8}
\definecolor{syntheticborder}{HTML}{C98B16}
\usetikzlibrary{arrows.meta,shapes.geometric}

\usepackage{inconsolata}

\usepackage{xcolor}
\usepackage{wrapfig}
\usepackage{tabularx}
\usepackage{caption}
\usepackage{tikz}
\usetikzlibrary{arrows.meta}
\usepackage{algorithm}
\usepackage{algpseudocode}
\usepackage[algo2e,ruled,vlined,linesnumbered,noend]{algorithm2e}
\usepackage{amsmath}
\usepackage{amssymb}\usepackage{colortbl}
\usepackage{multirow}
\usepackage{enumitem}
\newif\ifreviewcolors
\reviewcolorsfalse       %

\newcommand{\method}{\texorpdfstring{\textbf{\textsc{RCM}}}{RCM}}

\newcommand{\yes}{\textcolor{debategreen}{\(\checkmark\)}}
\newcommand{\no}{\textcolor{red!70!black}{\(\times\)}}
\newcommand{\partialmark}{\textcolor{openorange}{\(\circ\)}}

\title{A Responsibility-Centered Coordination Mechanism for Decentralized Multi-Agent Systems}

\author{Thi-Nhung Nguyen, Viet-Thanh Pham, Linhao Luo, \textbf{Thuy-Trang Vu} and \textbf{Dinh Phung} \\
  \textsuperscript{1} Department of Data Science \& AI, Monash University \\
  \texttt{\{nhung.thinguyen,thanh.pham1,linhao.luo1,trang.vu1,dinh.phung\}@monash.edu}
}

\begin{document}
\maketitle

\begin{abstract}
Centralized orchestration provides comprehensive multi-agent coordination but creates bottlenecks in scalability, adaptability, and collective performance. Decentralized approaches alleviate these bottlenecks by distributing coordination across agents, yet often leave responsibility for delegated dependencies and failure recovery unclear.
To address this gap, we propose \method{}, a \emph{responsibility-centered coordination mechanism} that
leverages \emph{commitments} to couple task delegation with explicit responsibilities of the involved parties.
Through recursive delegation, agents retain their upstream responsibilities while locally orchestrating downstream commitments, thereby preserving accountability across delegation chains, creating a network of \emph{local orchestration}. We operationalize this mechanism locally through responsibility-constrained decision-making and a \emph{responsibility-oriented communication protocol} (RCP), enabling agents to autonomously establish, manage, and terminate commitments. As agents act in parallel, their local actions continuously evolve the commitment graph, allowing global coordination to emerge and adapt toward a shared mission without centralized control. Across reasoning, coding, and executable agent environments demonstrate that \method{} outperforms strong centralized and decentralized coordination baselines by up to 18.7 points in task performance and 10.2 points in Responsibility, while improving failure and harm detection without task-specific training.
\end{abstract}

\section{Introduction}
\label{sec:introduction}

Coordination is essential for multi-agent systems (MAS) because it enables specialized agents to collaborate toward a shared mission. Centralized orchestration, ranging from predefined workflows to central orchestrators, has demonstrated effectiveness in coordinating complex tasks (e.g., \textsc{AutoGen} and \textsc{AgentVerse}) \citep{wu2023autogen,chen2024agentverse}. In these architectures, an orchestrator or prescribed control structure maintains a global view of the task, decomposes it into subtasks, assigns work to suitable agents, manages dependencies, monitors execution, and revises plans when failures occur.
However, centralized orchestration creates \emph{scalability bottlenecks} as coordination overhead and context grow, limits \emph{adaptability} because execution changes must pass through the central control point, and bounds collective \emph{performance} by the orchestrator's reasoning and inference capacity \citep{rana2000scalability,kim2026scaling,yang2026diversity}.

An alternative is decentralized coordination, which seeks scalable, adaptive, and resilient collaboration through peer-to-peer communication and local decision-making. Recent LLM-agent approaches distribute information exchange, collective reasoning, and adaptive task routing (e.g., \textsc{AgentNet}) \citep{yang2025agentnet,zhou2025reso,lee2026robust}, while local inference-time control guides individual agents during execution \citep{stolfo2025activation}. When an agent lacks the capabilities or resources to complete a task independently, it may act as the \emph{delegator} and delegate supporting work to another agent, the \emph{delegatee}. 
As delegation propagates across agents, downstream dependencies must be monitored, verified, and recovered from while preserving upstream commitments, which are functions traditionally handled by a central orchestrator. 
Existing decentralized coordination approaches may distribute task execution without explicitly assigning responsibility for managing these dependencies throughout execution. Consequently, these responsibilities may become \emph{fragmented} across agents, leaving ownership of downstream dependencies and responsibility for recovery unclear.

\begin{figure*}[t]
    \centering
    \includegraphics[width=0.95\textwidth]{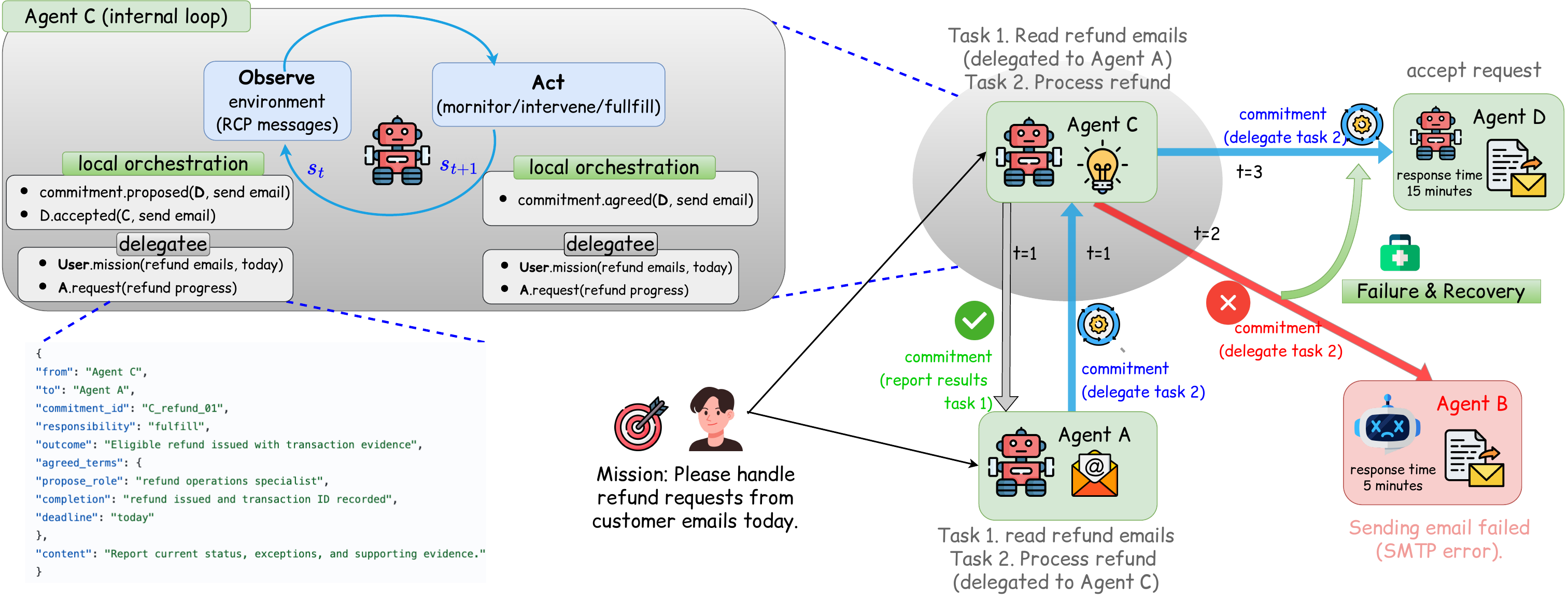}
    \caption{\textbf{Overview of \method{}.}
The right side shows global coordination through a dynamic network of commitments and local failure recovery. The left side illustrates each agent's internal observe--decide--act loop, through which local decisions collectively give rise to this global coordination.}

    \label{fig:mlo-overview}
\end{figure*}

To address this gap, we propose a \emph{\textbf{R}esponsibility-centered \textbf{C}oordination \textbf{M}echanism} (\method{}) that treats \emph{commitments} as the basic units of coordination. Each commitment combines a task delegation with explicit responsibilities for both parties. Upon opening a delegation, the delegatee is responsible for fulfilling the agreed task, while the delegator remains responsible for managing the resulting dependency until the commitment terminates. We refer to the delegator's ongoing responsibility as \emph{local orchestration}, as it involves functions traditionally performed by a central orchestrator, but only over the delegator's downstream commitments.
Through recursive delegation, agents retain their upstream commitments while locally orchestrating downstream dependencies, maintaining responsibility across delegation chains.

We operationalize this mechanism with \emph{responsibility-constrained decision-making} to guide each agent's actions according to its current responsibilities, and a \emph{responsibility-oriented communication protocol} (RCP) to support the establishment, management, and termination of commitments through inter-agent communication. Local orchestrators act in parallel, they observe changes in their local commitments and take appropriate actions (e.g., switching delegatees). These local actions continuously evolve the commitment graph, enabling global coordination to emerge and adapt toward the shared mission without a central orchestrator.

Our contributions are summarized as follows:
\begin{itemize}[noitemsep,topsep=2pt,parsep=2pt,partopsep=2pt]
    \item We propose a \emph{responsibility-centered coordination mechanism} (RCM) that uses commitments as units of coordination, preserving upstream responsibilities while managing downstream dependencies and failures across recursive delegation chains.
    \item We operationalize this mechanism at the local level through \emph{responsibility-constrained decision-making} and a \emph{responsibility-oriented communication protocol} (RCP), enabling global coordination to emerge from autonomous local decisions.
    \item We introduce responsibility-oriented evaluation metrics, including \emph{Communication}, \emph{Monitoring}, \emph{Intervention}, and \emph{Fulfillment} to assess how effectively agents uphold their responsibilities throughout execution and failure recovery.
    \item Across reasoning, coding, and executable agent environments, \method{} outperforms strong centralized and decentralized baselines, achieving a $10.2$-point gain in overall responsibility and up to an $18.7$-point gain in task performance, while improving failure and harmful-trajectory detection without task-specific training.
\end{itemize}

\section{Responsibility-centered Coordination Mechanism (RCM)}
\label{sec:method}

Consider a mission $q$ assigned to a team $\mathcal{A}=\{a_1,\ldots,a_n\}$ of autonomous agents with distributed capabilities. Each agent controls its own planner, tools, permissions, and execution state. Without a central orchestrator, the key challenge is how to coordinate autonomous agents while maintaining explicit responsibility as tasks are delegated and dependencies evolve.
As illustrated in Figure~\ref{fig:mlo-overview}, \method{} addresses this challenge by coordinating agents through commitments arising from task delegation. When $a_i$ delegates an outcome $g$ to $a_j$, a commitment $C_{i\rightarrow j}$ is formed, assigning local orchestration responsibility to $a_i$ and activating fulfilment responsibility for $a_j$ upon agreement. The formation and lifecycle of commitments establish and evolve coordination relationships among agents (Section~\ref{sec:commitments}).

Locally, each agent then acts autonomously according to its active responsibilities, using the \emph{responsibility-oriented communication protocol}  (RCP) to coordinate relevant changes with other agents.
As agents act in parallel, their local decisions evolve a network of interconnected commitments, allowing global coordination to emerge and adapt toward the shared mission without centralized control (Section~\ref{sec:local-decision}).

\subsection{Global Coordination through Commitments}
\label{sec:commitments}

Coordination across agents is established by linked commitments. Each commitment captures a delegated outcome, its lifecycle state, and the responsibilities that govern how its parties coordinate.

\paragraph{Commitment formation.}
When a delegator $a_i$ opens a delegation to a delegatee $a_j$, it creates a commitment record
\[
C_{i\rightarrow j}
=\langle d^{v}_{i\rightarrow j},s_C,\rho_C\rangle,
\]
where $d^{v}_{i\rightarrow j}$ is the delegation specification, $s_C$ tracks the commitment lifecycle from the initial request through resolution, and $\rho_C$ specifies the parties' state-dependent responsibilities.

The delegation specification describes the requested outcome and its terms:
\[
d^{v}_{i\rightarrow j}
=\langle\mathrm{id},v,a_i,a_j,g,\tau\rangle.
\]
Here, $\mathrm{id}$ distinguishes concurrent delegations, $v$ identifies the proposed or agreed version, and $g$ is the requested outcome. The terms $\tau$ specify the delegatee's task role, activation condition $\alpha$, completion criterion $\beta$, scope, permissions, resource budget, and timing. The parties may negotiate $g$ and $\tau$ while the commitment is in \textsc{Requested} state.
\begin{figure}[!t]
    \centering
    \includegraphics[width=\linewidth]{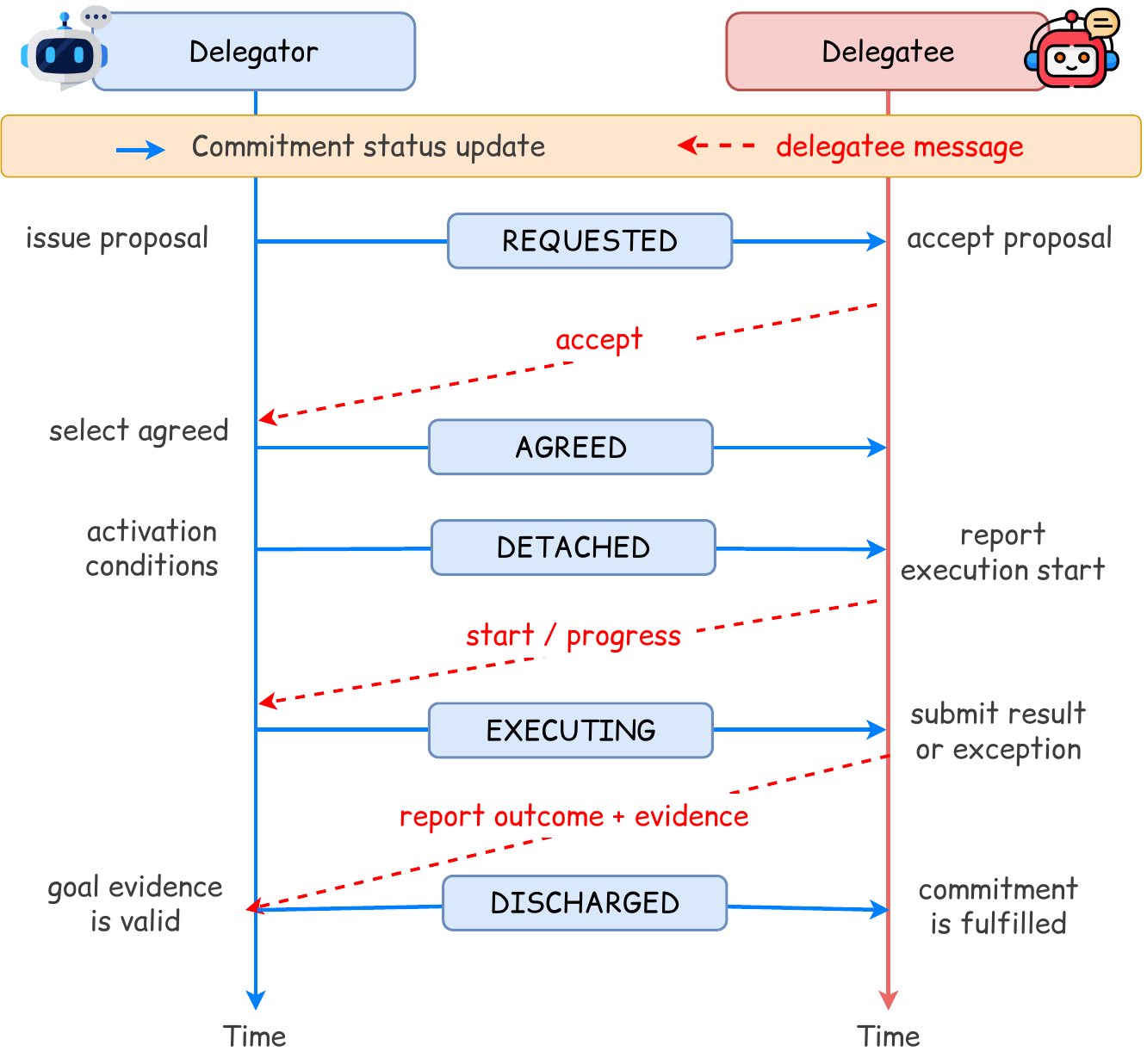}
    \caption{Example commitment handshake between a delegator and a delegatee. Blue arrows show commitment state updates; red dashed arrows show the delegatee's acceptance, progress reports, and outcome with completion evidence.}
    \label{fig:commitment-handshake}
\end{figure}

\paragraph{Commitment lifecycle}
The commitment lifecycle captures the delegation's execution status and the responsibilities active at each stage. Specifically, $s_C$ tracks the delegation state, while $\rho_C$ determines the corresponding active responsibilities. Upon opening a delegation, $a_i$ assumes \emph{local orchestration}: it manages the delegation, monitors its subsequent execution, and intervenes when failures or consequential changes threaten the requested outcome. This responsibility persists until the request is declined or the commitment terminates. While $s_C=\textsc{Requested}$, $\rho_C(a_j)=\varnothing$: the delegatee has no fulfilment responsibility. Once the parties agree on the delegation, $s_C$ becomes \textsc{Agreed}, and $\rho_C$ assigns \emph{fulfilment} to $a_j$. Fulfilment entails executing the agreed work, once $\alpha$ holds, and communicating consequential changes. These responsibilities are bounded by the agreed scope and authority, and revisions require renewed agreement. Verified completion or valid termination discharges the parties' active responsibilities. 
Figure~\ref{fig:commitment-handshake} illustrates the commitment lifecycle from its formation to verified completion.  See Appendix~\ref{app:commitment-record} for the complete list of possible commitment lifecycle statuses.

As agents recursively delegate work needed to fulfil their own commitments, the records form a dynamic coordination graph
$\mathcal G_t=\langle\mathcal A,\mathcal C_t,\mathcal R_t\rangle,$
where $\mathcal C_t$ contains live commitments and $\mathcal R_t$ records the support relations between them. The formation and state evolution of these linked commitments coordinate work across the team without requiring any agent to prescribe or observe a global execution plan.

\subsection{From Local Actions to Global Coordination}
\label{sec:local-decision}
\label{sec:commitment-graph}
\label{sec:agent-loop}

The preceding subsection describes how commitments establish coordination relationships among agents. We now explain how these relationships emerge and evolve from autonomous local actions governed by active responsibilities. To characterize decision making under limited visibility, we borrow the decentralized, partially observable perspective of a Dec-POMDP as a \emph{descriptive abstraction}. This formulation characterizes local observations and responsibility-constrained choices; it is not an agent-training procedure and does not require learning or solving a Dec-POMDP. We write
\[
\mathcal{M}_{\mathrm{MLO}}
=
\left\langle
\mathcal A,\mathcal S,
\{\mathcal O_i\}_{i\in\mathcal A},
\{\mathcal U_i\}_{i\in\mathcal A}
\right\rangle,
\]
where $\mathcal S$ is the global state space, while $\mathcal O_i$ and $\mathcal U_i$ are agent $a_i$'s local observation and action spaces.

The global state is
\[
s_t=\langle x_t,\mathcal G_t,\rho_t,M_t\rangle\in\mathcal S,
\]
where $x_t$ captures mission progress and relevant environmental conditions, $\mathcal G_t$ is the evolving commitment graph, and $\rho_t=\{\rho_C\}_{C\in\mathcal C_t}$ records the active responsibility allocations. The communication state $M_t=\{M_t^i\}_{i\in\mathcal A}$ contains messages exchanged among agents concerning live commitments, including delegation proposals, negotiation exchanges, execution reports, and updates.

No agent observes $s_t$ in full. Instead, agent $a_i$ receives the local observation
\[
o_t^i=
\langle x_t^i,\widehat{\mathcal G}_t^i,\rho_t^i,M_t^i\rangle
\in\mathcal O_i,
\]
where $x_t^i$ contains the execution progress and environmental conditions directly observable for the work it owns or performs; $\widehat{\mathcal G}_t^i$ contains its local view of the commitments to which it is a party and their support relations; $\widehat{\mathcal C}_t^i\subseteq\mathcal C_t$ contains the live commitments in that view; $\rho_t^i=\bigcup_{C\in\widehat{\mathcal C}_t^i}\rho_C(a_i)$ is its active responsibility set; and $M_t^i=\{m\in M_t\mid a_r(m)=a_i\}$ contains the messages addressed to $a_i$.

\begin{table}[!t]
\centering
\scriptsize
\setlength{\tabcolsep}{1.5pt}
\renewcommand{\arraystretch}{0.92}
\setlength{\aboverulesep}{0.6pt}
\setlength{\belowrulesep}{0.6pt}
\begin{tabularx}{\columnwidth}{@{}>{\raggedright\arraybackslash}p{0.30\columnwidth}>{\raggedright\arraybackslash}X@{}}
\toprule
\textbf{RCP operation} & \textbf{Initiator and semantics} \\
\midrule
\texttt{delegate.propose} & $a_i\to a_j$: open a delegation; record the $g,\tau$ proposal as \textsc{Requested}. \\
\texttt{delegate.accept}/\newline\texttt{decline} & $a_j\to a_i$: accept or decline the proposal version; acceptance establishes commitment edge $C$ in \textsc{Agreed}. \\
\texttt{commitment.query}/\newline\texttt{communicate} & $a_i\leftrightarrow a_j$: exchange questions, clarifications, information, or intervention requests within scope; correlate responses by \texttt{reply\_to}. \\
\texttt{commitment.notify} & $a_i\leftrightarrow a_j$: report consequential changes; $a_j$ also reports start, progress, partial results, and exceptions. \\
\texttt{outcome.submit}/\newline\texttt{verify} & $a_j\to a_i$: submit outcome evidence; $a_i$ verifies it against $\beta$ before discharge. \\
\texttt{commitment.revise} & $a_i\leftrightarrow a_j$: counter or revise $g,\tau$; changes take effect only after agreement on the new version. \\
\texttt{commitment.update}/\newline\texttt{terminate} & Local orchestration by $a_i$: update $s_C$ on edge $C$ using validated conditions and evidence; termination requires authority under $\tau$. \\
\bottomrule
\end{tabularx}
\caption{Core RCP operations between delegator $a_i$ and delegatee $a_j$. Messages inform the recipient's decisions under its active responsibility in $\rho_C$; lifecycle updates are guarded changes to commitment edges.}
\label{tab:rcp-operations}
\end{table}

\paragraph{Responsibility-constrained actions.}
Given this local view, agent $a_i$ selects actions according to its active responsibilities. The local action space $\mathcal U_i$ comprises three classes of actions: (i) \emph{fulfil an upstream commitment}, by performing the agreed task and reporting its outcome; (ii) \emph{create a downstream commitment}, by delegating required supporting work; and (iii) \emph{intervene in or update a downstream commitment}, by monitoring, verifying, revising, or recovering delegated work. We denote these classes by
\[
\mathcal U_i
=\mathcal U_i^{\mathrm{fulfil}}
\cup\mathcal U_i^{\mathrm{create}}
\cup\mathcal U_i^{\mathrm{update}}.
\]
The agent's local action--observation history is $h_t^i=(o_0^i,u_0^i,\ldots,u_{t-1}^i,o_t^i)$. Its own decision policy $\pi_i$ selects an action subject to the responsibility constraint:
\begin{equation}
\label{eq:responsibility-constrained-action}
\begin{gathered}
u_t^i\sim\pi_i(\cdot\mid h_t^i)\\
\text{subject to}\quad c_{\mathrm{resp}}(u_t^i;o_t^i)=0.
\end{gathered}
\end{equation}
Here, $\pi_i$ denotes the agent's autonomous decision policy, not a policy trained by \method{}. The constraint $c_{\mathrm{resp}}=0$ only when the selected action is consistent with the agent's active responsibilities and permitted by the relevant commitment state and applicable delegation terms; otherwise $c_{\mathrm{resp}}>0$. In implementation, \method{} applies a set of rules based on the agent's role, the commitment lifecycle, and the agreed terms. For example, execution while the commitment is still \textsc{Requested}, or an attempt to move a \textsc{Discharged} commitment back to \textsc{Executing}, violates the lifecycle and yields $c_{\mathrm{resp}}>0$.

\paragraph{Communication and commitment-state updates.}
When taking a local action $u_t^i\in\mathcal U_i$, agent $a_i$ may need to coordinate with its delegator or delegatee to fulfil an upstream commitment or create and manage downstream commitments. To support these responsibility-driven interactions, agents communicate through the \emph{responsibility-oriented communication protocol} (RCP), which standardizes message operations for delegation negotiation, information exchange, execution reporting, outcome verification, and guarded commitment updates (Table~\ref{tab:rcp-operations}).

RCP messages in $M_t$, together with evidence and results of task execution, inform subsequent local decisions and may trigger transitions in commitment state $s_C$. Each transition is applied only after its conditions and evidence are verified. In particular, a delegatee reports consequential changes to its upstream delegator, enabling the latter to monitor or intervene in the corresponding downstream commitment. Updates are exchanged only with relevant counterparts. RCP can run over existing messaging systems, while agents retain their own planners and tools; see Appendix~\ref{app:rcp-details} for message processing and delivery.

\paragraph{Dynamic Emergence and Evolution of the Commitment Graph.}
As agents fulfil their upstream commitments, they may recursively delegate supporting work, establishing chains of downstream commitments. These commitments and their support relations collectively form the commitment graph $\mathcal G_t$.
Agents observe changes in their local commitments and take necessary actions accordingly. For example, when a downstream commitment becomes \textsc{Violated}, the delegator closes it and may establish a replacement commitment with another delegatee, while unaffected work continues independently. When a commitment terminates, its dependent supporting commitments are closed or released under their agreed terms. As agents act in parallel, these local actions continuously evolve $\mathcal G_t$, giving rise to a dynamic execution workflow toward the shared mission $q$ without a central orchestrator. Algorithm~\ref{alg:recursive-coordination} illustrates how agents recursively fulfil upstream commitments and locally orchestrate downstream commitments.

\begin{algorithm2e}[!t]
\tiny\raggedright
\SetAlFnt{\tiny\raggedright}
\SetAlCapFnt{\small}
\SetAlCapNameFnt{\small}
\SetArgSty{textnormal}
\SetFuncSty{textsc}
\SetInd{0.3em}{0.7em}
\DontPrintSemicolon
\SetKwInput{Require}{Require}
\SetKwFunction{Coordinate}{Coordinate}
\SetKwProg{CoordProc}{procedure}{:}{}

\Require{Mission $q$, agents $\mathcal A$,
initial commitments $\mathcal C_0(q)$,
delegation limits $K,L$}

\ForEach{$C\in\mathcal C_0(q)$ with $s_C=\textsc{Agreed}$
\textbf{in parallel}}{
    Start \Coordinate{$\operatorname{delegatee}(C),C$}\;
}

\CoordProc{\Coordinate{$a,C^{\uparrow}$}}{
    \While{$C^{\uparrow}$ is live}{
        Observe $o_t^a$; verify commitment-state changes\;
        Select $u_t^a\in\mathcal U_a$ such that
        $c_{\mathrm{resp}}(u_t^a;o_t^a)=0$\;

        \uIf{$u_t^a\in\mathcal U_a^{\mathrm{fulfil}}$}{
            Execute $g^{\uparrow}$ when active;
            report results and evidence via RCP\;
        }
        \uElseIf{$u_t^a\in\mathcal U_a^{\mathrm{create}}$}{
            Open downstream commitments in \textsc{Requested};
            negotiate via RCP within $(K,L)$ limits\;
            Record support links\;
        }
        \ElseIf{$u_t^a\in\mathcal U_a^{\mathrm{update}}$}{
            Monitor and update downstream commitments\;
            Close \textsc{Violated} commitments;
            reassess support and replacement if needed\;
        }

        \ForEach{newly \textsc{Agreed} $C^{\downarrow}$
        \textbf{in parallel}}{
            Start \Coordinate{
            $\operatorname{delegatee}(C^{\downarrow}),C^{\downarrow}$}\;
        }
        Notify relevant counterparts of consequential changes via RCP\;
    }
    Close pending proposals; release dependent commitments
    under their terms\;
    Propagate termination; prune closed records and links\;
}
\caption{Recursive coordination through responsibility-constrained local actions.}
\label{alg:recursive-coordination}
\end{algorithm2e}

\section{Experimental Setup}
\label{sec:setup}

\begin{table*}[t]
\centering
\scriptsize
\setlength{\tabcolsep}{2.0pt}
\renewcommand{\arraystretch}{0.90}
\resizebox{\textwidth}{!}{
\begin{tabular}{clccc@{\hspace{2pt}\vrule width 0.35pt\hspace{2pt}}cccc@{\hspace{2pt}\vrule width 0.35pt\hspace{2pt}}cccc}
\toprule
\multirow{3}{*}{\textbf{Workflow}} & \multirow{3}{*}{\textbf{Method}} & \multirow{3}{*}{\shortstack{\textbf{Dynamic}\\\textbf{workflow}}} & \multirow{3}{*}{\shortstack{\textbf{Responsi-}\\\textbf{bility}}} & \multirow{3}{*}{\shortstack{\textbf{Distributed}\\\textbf{M/R}}} & \multicolumn{4}{c}{\textbf{Qwen3-4B}} & \multicolumn{4}{c}{\textbf{GPT-5.4}} \\[-1.5pt]
\cmidrule(lr){6-9}\cmidrule(lr){10-13}
& & & & & \multicolumn{2}{c}{\textbf{Non-Agentic}} & \multicolumn{2}{c}{\textbf{Agentic}} & \multicolumn{2}{c}{\textbf{Non-Agentic}} & \multicolumn{2}{c}{\textbf{Agentic}} \\[-1.5pt]
\cmidrule(lr){6-7}\cmidrule(lr){8-9}\cmidrule(lr){10-11}\cmidrule(lr){12-13}
& & & & & \textbf{AIME25} & \textbf{HumanEval} & \textbf{Claw-Eval} & \textbf{Claw-Live} & \textbf{AIME25} & \textbf{HumanEval} & \textbf{Claw-Eval} & \textbf{Claw-Live} \\[-1pt]
\midrule
\textbf{Broadcast} & MAS-Broadcast & \no & Unspecified & \no & 30.0{\tiny$\pm$3.3} & 83.3{\tiny$\pm$2.2} & 35.7{\tiny$\pm$1.9} & 40.2{\tiny$\pm$3.7} & 43.3{\tiny$\pm$8.4} & 93.9{\tiny$\pm$1.5} & 50.5{\tiny$\pm$1.4} & 64.4{\tiny$\pm$2.7} \\
\cmidrule(lr){1-13}
\multirow{5}{*}{\textbf{Designed}} & MachineSoM & \no & Debate & \no & 16.7{\tiny$\pm$5.1} & 87.9{\tiny$\pm$0.9} & 41.6{\tiny$\pm$1.4} & 56.3{\tiny$\pm$1.5} & \cellcolor{debategreen!25}\textbf{66.7}{\tiny$\pm$0.0} & 90.9{\tiny$\pm$4.0} & 44.2{\tiny$\pm$1.5} & 51.5{\tiny$\pm$3.1} \\
 & Sparse MAD & \no & Debate & \no & 6.7{\tiny$\pm$3.3} & 81.8{\tiny$\pm$2.3} & 43.0{\tiny$\pm$0.5} & \cellcolor{debategreen!12}\underline{59.5}{\tiny$\pm$0.8} & 60.0{\tiny$\pm$10.2} & \cellcolor{debategreen!25}\textbf{98.5}{\tiny$\pm$0.0} & 46.9{\tiny$\pm$0.9} & 60.0{\tiny$\pm$1.1} \\
 & MetaGPT & \no & Role & \no & \cellcolor{debategreen!12}\underline{36.7}{\tiny$\pm$8.8} & 86.4{\tiny$\pm$1.5} & 40.5{\tiny$\pm$1.9} & 52.4{\tiny$\pm$1.0} & \cellcolor{debategreen!12}\underline{63.3}{\tiny$\pm$5.1} & 89.4{\tiny$\pm$3.5} & 51.4{\tiny$\pm$2.6} & 57.6{\tiny$\pm$2.4} \\
 & AutoGen & \no & Role & \no & 26.7{\tiny$\pm$1.9} & \cellcolor{debategreen!12}\underline{90.9}{\tiny$\pm$1.8} & 36.0{\tiny$\pm$0.3} & 35.1{\tiny$\pm$3.0} & 53.3{\tiny$\pm$8.4} & 93.9{\tiny$\pm$0.9} & 54.0{\tiny$\pm$1.4} & 70.8{\tiny$\pm$1.5} \\
 & AgentVerse & \partialmark & Role & \no & 26.7{\tiny$\pm$5.6} & 81.8{\tiny$\pm$3.2} & \cellcolor{debategreen!12}\underline{44.3}{\tiny$\pm$1.4} & 57.3{\tiny$\pm$0.8} & \cellcolor{debategreen!12}\underline{63.3}{\tiny$\pm$0.0} & 87.9{\tiny$\pm$3.0} & \cellcolor{debategreen!12}\underline{66.5}{\tiny$\pm$0.3} & \cellcolor{debategreen!12}\underline{71.9}{\tiny$\pm$1.3} \\
\cmidrule(lr){1-13}
\multirow{4}{*}{\textbf{Learned}} & AFlow & \no & Implicit & \no & 23.3{\tiny$\pm$5.1} & \_ & 40.6{\tiny$\pm$0.7} & 53.9{\tiny$\pm$1.0} & \cellcolor{debategreen!25}\textbf{66.7}{\tiny$\pm$1.9} & 95.5{\tiny$\pm$0.9} & 48.2{\tiny$\pm$0.9} & 55.5{\tiny$\pm$1.7} \\
 & GPTSwarm & \no & Implicit & \no & 30.0{\tiny$\pm$5.1} & 81.8{\tiny$\pm$1.5} & 40.2{\tiny$\pm$0.8} & 46.5{\tiny$\pm$1.7} & 56.7{\tiny$\pm$11.5} & 95.5{\tiny$\pm$0.9} & 50.0{\tiny$\pm$1.2} & 59.4{\tiny$\pm$3.1} \\
 & AgentNet & \yes & Execution & \no & 13.3{\tiny$\pm$3.3} & 86.4{\tiny$\pm$1.5} & 36.4{\tiny$\pm$1.8} & 48.5{\tiny$\pm$0.2} & 53.3{\tiny$\pm$6.7} & 93.9{\tiny$\pm$0.9} & 49.1{\tiny$\pm$0.4} & 50.0{\tiny$\pm$3.0} \\
 & AgentPrune & \no & Implicit & \no & 34.6{\tiny$\pm$2.0} & 84.8{\tiny$\pm$2.3} & 38.6{\tiny$\pm$1.6} & 44.8{\tiny$\pm$2.0} & 50.0{\tiny$\pm$7.7} & 95.5{\tiny$\pm$0.9} & 51.4{\tiny$\pm$0.7} & 61.7{\tiny$\pm$1.2} \\
\cmidrule(lr){1-13}
\textbf{Emergent} & \textbf{\method{}} & \yes & \textbf{Commitment} & \yes & \cellcolor{debategreen!25}\textbf{50.0}{\tiny$\pm$5.8} & \cellcolor{debategreen!25}\textbf{91.9}{\tiny$\pm$0.0} & \cellcolor{debategreen!25}\textbf{63.0}{\tiny$\pm$2.2} & \cellcolor{debategreen!25}\textbf{61.4}{\tiny$\pm$1.6} & \cellcolor{debategreen!25}\textbf{66.7}{\tiny$\pm$0.0} & \cellcolor{debategreen!12}\underline{97.0}{\tiny$\pm$0.9} & \cellcolor{debategreen!25}\textbf{72.8}{\tiny$\pm$1.3} & \cellcolor{debategreen!25}\textbf{79.4}{\tiny$\pm$0.3} \\
\bottomrule
\end{tabular}}
\caption{Performance across tasks, environments, and executor models. Dynamic workflow indicates whether task relationships can change during execution. Responsibility is unspecified under broadcast, induced through debate or role assignment, implicit in workflow structure, execution-based, or explicit through commitment. Distributed M/R denotes distributed monitoring and recovery. \yes, \partialmark, and \no indicate full, partial, and no support.}
\label{tab:illustrative-e2e}
\end{table*}

\paragraph{Evaluation Data:}
\label{sec:tasks}
\textsc{AIME25} \citep{aime25} and \textsc{HumanEval} \citep{chen2021humaneval} test whether coordination preserves mathematical reasoning and functional code generation, respectively. \textsc{Claw-Eval} \citep{ye2026claweval} evaluates multi-service tasks under controlled failures, and \textsc{Claw-Eval-Live} \citep{li2026clawevallive} evaluates changing real-world workflows. In the latter two, we randomly partition the available tools across all agents in each team, creating a distributed tool environment that requires delegation and communication. All methods use identical task instances, tool allocations, permissions, prompts, and test-time budgets. Appendix~\ref{app:benchmark-details} gives benchmark details.

\paragraph{Metrics:} We evaluate \method{} along two dimensions: task capability and responsibility. 
\begin{itemize}[noitemsep,topsep=2pt,parsep=2pt,partopsep=2pt,leftmargin=*]
\item 
\textbf{Task capability}is measured using accuracy (AIME25), Pass@1 (\textsc{HumanEval}), score and completion (\textsc{Claw-Eval}), and score (\textsc{Claw-Eval-Live}). 

\item \textbf{Responsibility} measures how effectively agents handle failures during execution. We define an \emph{environmental failure} as an explicit unsuccessful response from a tool or the execution environment, and define four metrics: (i) \emph{Communicate} is the fraction of environmental failures subsequently reported by an agent; (ii) \emph{Monitor} is the fraction of environmental failures followed by a relevant change in action; (iii) \emph{Intervene} is the fraction of such changed actions that successfully resolve or bypass the corresponding failure; (iv) \emph{Fulfillment} is the mean official \textsc{Claw-Eval} completion score. Their equal-weight mean gives the \emph{Responsibility score}. Failure signals and deduplication are specified in Appendix~\ref{app:failure-signals}. 
\end{itemize}

\paragraph{Baselines:}
We consider three families of multi-agent baselines:
\begin{itemize}[noitemsep,topsep=2pt,parsep=2pt,partopsep=2pt,leftmargin=*]
\item \emph{\textbf{Decentralized}}: \textsc{MAS-Broadcast}, \textsc{MachineSoM}, \textsc{Sparse MAD}, and \textsc{AgentNet} \citep{zhang2024machinesom,li2024sparsemad,yang2025agentnet};
\item \emph{\textbf{Hand-designed}}: \textsc{MetaGPT}, \textsc{AutoGen}, and \textsc{AgentVerse} \citep{hong2024metagpt,wu2023autogen,chen2024agentverse};
\item \emph{\textbf{Optimized}}: \textsc{AFlow}, \textsc{GPTSwarm}, and \textsc{AgentPrune} \citep{zhang2025aflow,zhuge2024gptswarm,zhang2025agentprune}.
On Who\&When, we include Aegis attribution models \citep{kong2026aegis} and training-free LLMs. On ATBench-C, we include guard models \citep{meta2024llamaguard3,meta2025llamaguard4,zhao2025qwen3guard,chen2025shieldagent,liu2026agentdog} and general-purpose LLMs. 
\end{itemize}
Baseline and implementation details are in Appendices~\ref{app:baseline-details} and~\ref{app:implementation-details}, respectively.

\section{Results}
\label{sec:results}

Our evaluation first examines whether \method{} enables robust coordination across tasks and teams, then analyzes how explicit responsibilities support reliable coordination. 

\subsection{Robustness and Autonomous Adaptation}
\paragraph{Main results.}

Table~\ref{tab:illustrative-e2e} reports overall performance on reasoning and coding benchmarks that do not require agent interaction and on executable agentic tasks. For \textsc{Claw-Eval} and \textsc{Claw-Eval-Live}, tools are partitioned across agents so that no single agent can complete the task independently; success therefore requires delegation and communication rather than stronger individual execution alone.
\begin{figure}[t]
\centering
\includegraphics[width=\linewidth]{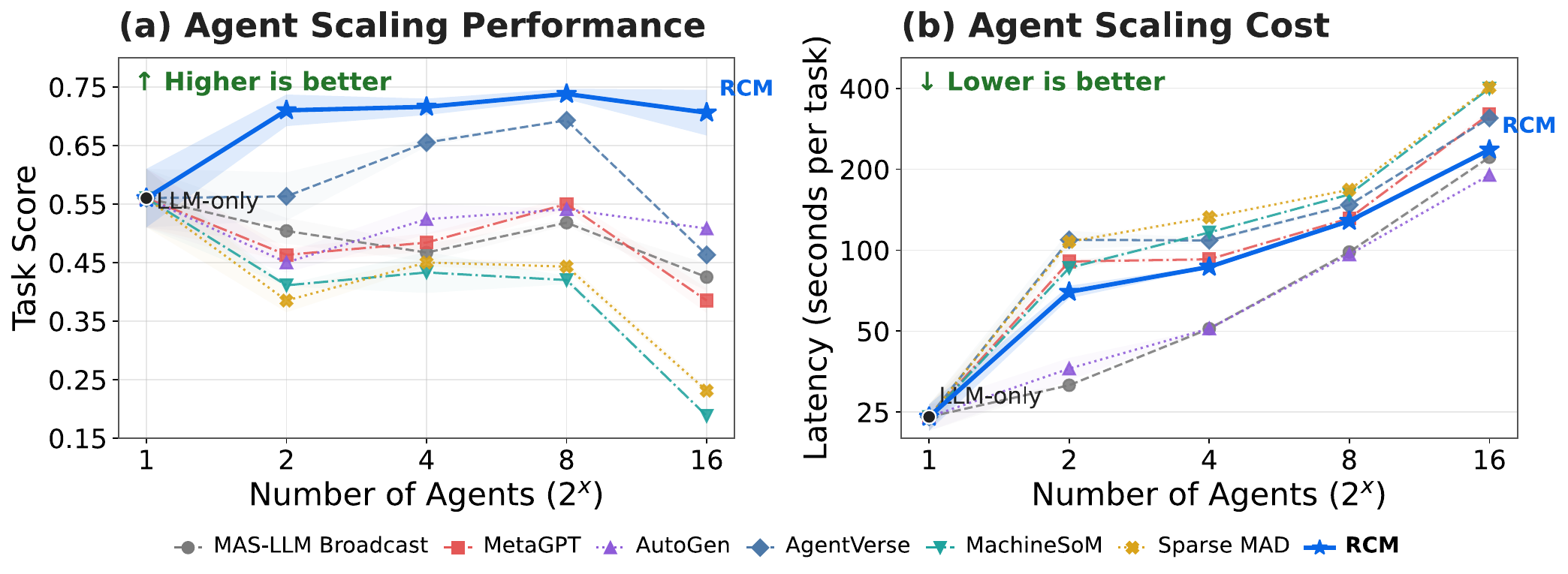}
\caption{Agent scaling on \textsc{Claw-Eval}.}
\label{fig:agent-scaling}
\end{figure}

Across Qwen3-4B experiments, \method{} achieves the best result in all four columns. With GPT-5.4, it leads both agentic benchmarks, ties the best AIME25 result, and remains within 1.5 points of the best HumanEval result. \emph{The advantage is largest when coordination is essential.} Relative to the strongest baseline in each column, \method{} improves \textsc{Claw-Eval} by 18.7 points with Qwen3-4B and 6.3 points with GPT-5.4, and \textsc{Claw-Eval-Live} by 1.9 and 7.5 points, respectively. Overall, explicit responsibility for each commitment enables robust and autonomous coordination without sacrificing the executor's reasoning and coding capability.

\paragraph{Adaptation to changing agent teams.}
Figure~\ref{fig:agent-scaling} tests whether this robustness persists as the team changes. We increase the team from 1 to 16 agents and divide a fixed tool set more finely among them, progressively increasing the need for distributed coordination. \emph{\method{} maintains the strongest task performance throughout this scaling process.} Several baselines are faster with small teams, but their latency grows more sharply as agents are added, while \method{} becomes more efficient at larger team sizes. These results show that coordination centered on responsibility can adapt to new team configurations without requiring a separate workflow for each configuration.

\subsection{How Responsibility Shapes Coordination}
\paragraph{Responsibility during execution.}
Table~\ref{tab:illustrative-responsibility} evaluates whether agents communicate consequential failures, monitor the affected commitment, intervene effectively, and still fulfill the delegated outcome. \emph{Several baselines communicate and monitor failures reliably but fail to convert detection into successful response.} For example, GPTSwarm reaches 95.2\% Communicate and 95.0\% Monitor, yet only 32.5\% Intervene and 37.8\% Fulfillment.

\emph{In contrast, \method{} performs consistently across the full responsibility chain.} It reaches 100.0\% Communicate, 99.0\% Monitor, 67.0\% Intervene, and 65.4\% Fulfillment, yielding a Responsibility score of 82.8\%, 10.2 points above the strongest baseline. \emph{Higher responsibility is associated with stronger task performance.} Across methods, the Responsibility score exhibits a strong positive linear correlation with the corresponding Claw-Eval task score in 
Table~\ref{tab:illustrative-e2e}.
\begin{table}[!t]
\centering
\scriptsize
\setlength{\tabcolsep}{1.3pt}
\renewcommand{\arraystretch}{0.90}
\setlength{\aboverulesep}{0.6pt}
\setlength{\belowrulesep}{0.6pt}
\begin{tabularx}{\columnwidth}{@{}>{\raggedright\arraybackslash}Xccccc@{}}
\toprule
\textbf{Method} & \textbf{Comm.}$\uparrow$ & \textbf{Mon.}$\uparrow$ & \textbf{Intv.}$\uparrow$ & \textbf{Ful.}$\uparrow$ & \textbf{Resp.}$\uparrow$ \\
\midrule
\rowcolor{gray!12}\multicolumn{6}{c}{\textit{Broadcast}} \\
MAS-Broadcast & 82.9{\tiny$\pm$5.0} & 89.7{\tiny$\pm$4.7} & 44.8{\tiny$\pm$5.2} & 38.1{\tiny$\pm$1.7} & 63.9{\tiny$\pm$2.7} \\
\midrule
\rowcolor{gray!12}\multicolumn{6}{c}{\textit{Designed}} \\
MachineSoM & 94.7{\tiny$\pm$6.0} & 77.8{\tiny$\pm$4.8} & 13.9{\tiny$\pm$5.1} & 30.2{\tiny$\pm$1.5} & 54.1{\tiny$\pm$1.5} \\
Sparse MAD & 93.9{\tiny$\pm$0.4} & 77.4{\tiny$\pm$1.3} & 12.9{\tiny$\pm$1.4} & 33.7{\tiny$\pm$1.1} & 54.5{\tiny$\pm$0.9} \\
MetaGPT & 75.4{\tiny$\pm$10.0} & 83.7{\tiny$\pm$7.7} & 51.2{\tiny$\pm$10.1} & 39.8{\tiny$\pm$2.9} & 62.5{\tiny$\pm$5.2} \\
AutoGen & 73.7{\tiny$\pm$8.8} & 71.4{\tiny$\pm$4.7} & \cellcolor{debategreen!12}\underline{64.3}{\tiny$\pm$13.4} & 42.8{\tiny$\pm$1.7} & 63.0{\tiny$\pm$2.9} \\
AgentVerse & \cellcolor{debategreen!12}\underline{95.8}{\tiny$\pm$0.8} & \cellcolor{debategreen!12}\underline{95.7}{\tiny$\pm$0.8} & 40.6{\tiny$\pm$4.2} & \cellcolor{debategreen!12}\underline{58.4}{\tiny$\pm$0.2} & \cellcolor{debategreen!12}\underline{72.6}{\tiny$\pm$1.2} \\
\midrule
\rowcolor{gray!12}\multicolumn{6}{c}{\textit{Learned}} \\
AFlow & 92.7{\tiny$\pm$0.8} & 92.1{\tiny$\pm$1.2} & 18.4{\tiny$\pm$7.2} & 35.2{\tiny$\pm$0.9} & 59.6{\tiny$\pm$2.1} \\
GPTSwarm & 95.2{\tiny$\pm$8.4} & 95.0{\tiny$\pm$2.3} & 32.5{\tiny$\pm$4.2} & 37.8{\tiny$\pm$1.0} & 65.1{\tiny$\pm$3.4} \\
AgentNet & 92.7{\tiny$\pm$5.3} & 84.2{\tiny$\pm$4.0} & 42.1{\tiny$\pm$3.4} & 36.9{\tiny$\pm$0.9} & 64.0{\tiny$\pm$1.6} \\
AgentPrune & 93.8{\tiny$\pm$2.2} & 93.3{\tiny$\pm$4.0} & 43.3{\tiny$\pm$13.3} & 39.2{\tiny$\pm$0.6} & 67.4{\tiny$\pm$4.3} \\
\midrule
\rowcolor{gray!12}\multicolumn{6}{c}{\textit{Emergent}} \\
\rowcolor{debategreen!25}
\textbf{\method{}} & \textbf{100.0}{\tiny$\pm$0.0} & \textbf{99.0}{\tiny$\pm$0.7} & \textbf{67.0}{\tiny$\pm$1.0} & \textbf{65.4}{\tiny$\pm$0.4} & \textbf{82.8}{\tiny$\pm$0.0} \\
\bottomrule
\end{tabularx}
\caption{Responsibility metrics on \textsc{Claw-Eval} with GPT-5.4. Resp. is the equal-weight mean of Comm., Mon., Intv., and Ful.}
\label{tab:illustrative-responsibility}
\end{table}

\paragraph{Local Orchestration for Safe and Resilient Coordination.}
To evaluate the effectiveness of monitoring in local orchestration, we apply \method{}'s local-orchestration mechanism to detecting execution failures and harmful trajectories on Who\&When and ATBench-C, respectively. Who\&When evaluates whether the monitor identifies the responsible agent and decisive error step, while ATBench-C assesses its ability to detect harmful trajectories before an intervention is accepted.

\begin{table}[t]
\centering
\scriptsize
\setlength{\tabcolsep}{1.5pt}
\renewcommand{\arraystretch}{0.90}
\setlength{\aboverulesep}{0.6pt}
\setlength{\belowrulesep}{0.6pt}
\begin{tabularx}{\columnwidth}{@{}>{\raggedright\arraybackslash}Xcccc@{}}
\toprule
\textbf{Model} & \textbf{Pair} & \textbf{Agent} & \textbf{Error} & \textbf{Avg.} \\
\midrule
\rowcolor{gray!12}\multicolumn{5}{c}{\textit{Trained baselines (Trained \yes)}} \\
DCL (Aegis) & 1.60 & 8.40 & 14.67 & 8.22 \\
Aegis-SFT (Qwen2.5-14B) & 4.03 & 51.14 & 9.87 & 21.68 \\
Aegis-GRPO (Qwen2.5-14B) & 2.45 & 54.43 & 4.15 & 20.34 \\
Qwen3-8B-Thinking + GRPO & \underline{8.10} & 53.12 & 11.25 & 24.16 \\
\midrule
\rowcolor{gray!12}\multicolumn{5}{c}{\textit{Training-free baselines (Trained \no)}} \\
Qwen3-8B-Thinking & 1.95 & 37.91 & 4.65 & 14.84 \\
gpt-oss-120b & 8.09 & 51.56 & 14.75 & 24.80 \\
o3 & 7.41 & 53.10 & \underline{14.88} & \underline{25.13} \\
Gemini-2.5-Flash & 7.32 & \underline{55.56} & 11.94 & 24.94 \\
Claude-Sonnet-4 & 6.77 & 44.76 & 13.33 & 21.62 \\
\midrule
\rowcolor{cyan!12}
\textbf{\method{} (Qwen3-4B)} & 7.07 & 54.89 & 10.87 & 24.28 \\
\rowcolor{cyan!12}
\textbf{\method{} (GPT-5.4)} & \textbf{14.67} & \textbf{66.30} & \textbf{23.91} & \textbf{34.96} \\
\bottomrule
\end{tabularx}
\caption{Failure attribution on Who\&When under the Aegis protocol. Baseline results are from \citet{kong2026aegis}. Avg. is the mean of Pair, Agent, and Error micro-F1.}
\label{tab:failure-attribution}
\end{table}

\begin{table}[t]
\centering
\scriptsize
\setlength{\tabcolsep}{1.5pt}
\renewcommand{\arraystretch}{0.90}
\setlength{\aboverulesep}{0.6pt}
\setlength{\belowrulesep}{0.6pt}
\begin{tabularx}{\columnwidth}{@{}>{\raggedright\arraybackslash}Xcccc@{}}
\toprule
\textbf{Model} & \textbf{Acc.} & \textbf{Prec.} & \textbf{Rec.} & \textbf{F1} \\
\midrule
\rowcolor{gray!12}\multicolumn{5}{c}{\textit{Guard models (Trained \yes)}} \\
LlamaGuard3-8B & 53.1 & \textbf{85.7} & 3.8 & 7.3 \\
LlamaGuard4-12B & 58.1 & 63.8 & 30.9 & 41.7 \\
Qwen3-Guard & 51.5 & 40.0 & 0.4 & 0.8 \\
ShieldAgent & 62.5 & 58.0 & 81.4 & 67.7 \\
AgentDoG-Qwen3-4B & 64.0 & 59.2 & \textbf{88.9} & \underline{71.1} \\
\midrule
\rowcolor{gray!12}\multicolumn{5}{c}{\textit{Open-source models (Trained \no)}} \\
Qwen3-4B-Instruct-2507 & 55.7 & \underline{77.6} & 15.3 & 25.5 \\
Qwen3-235B-A22B-Instruct-2507 & 59.2 & 58.2 & 63.8 & 60.8 \\
Qwen3.5-397B-A17B & \underline{66.8} & 65.5 & 70.2 & 67.8 \\
\midrule
\rowcolor{cyan!12}
\textbf{\method{} (Qwen3-4B-Instruct-2507)} & \textbf{71.20} & 67.33 & \underline{81.69} & \textbf{73.82} \\
\bottomrule
\end{tabularx}
\caption{Harmful-trajectory detection on ATBench-C. Baseline results are from \citet{li2026atbench}.}
\label{tab:atbench}
\vspace{-1em}
\end{table}

On Who\&When (Table~\ref{tab:failure-attribution}), \method{} with GPT-5.4 achieves an average F1 of 34.96, outperforming the strongest selected baseline (o3) by 9.83 points. On ATBench-C (Table~\ref{tab:atbench}), \method{} with Qwen3-4B-Instruct-2507 achieves an F1 of 73.82, surpassing the strongest selected guard model by 2.72 points. Notably, these results are achieved without task-specific training, suggesting that local orchestration can enable agents to act autonomously and safely while remaining resilient to failures.

\paragraph{Component contributions.}
Table~\ref{tab:illustrative-ablation} separates the mechanisms that make responsibility operational. Restricting initial task receivers to a single agent, whether relevance-based or random, degrades performance compared to assigning the task to all agents initially.
Removing any component further degrades performance, confirming that each contributes to the overall effectiveness of \method{}.

\begin{table}[t]
\centering
\scriptsize
\setlength{\tabcolsep}{2.2pt}
\renewcommand{\arraystretch}{0.90}
\setlength{\aboverulesep}{0.6pt}
\setlength{\belowrulesep}{0.6pt}
\begin{tabularx}{\columnwidth}{@{}>{\raggedright\arraybackslash}Xcc@{}}
\toprule
\textbf{Variant} & \textbf{Task} $\uparrow$ & \textbf{Resp.} $\uparrow$ \\
\midrule
\rowcolor{cyan!12}
\textbf{Full \method{}} & \textbf{72.8}{\tiny$\pm$1.3} & \textbf{82.8}{\tiny$\pm$0.0} \\
Relevant-1 initial receiver & 61.5{\tiny$\pm$1.6} & 73.9{\tiny$\pm$5.0} \\
Random-1 initial receiver & 59.1{\tiny$\pm$1.4} & 67.5{\tiny$\pm$3.6} \\
\midrule
Without explicit delegator responsibility & 60.4{\tiny$\pm$1.7} & 68.5{\tiny$\pm$1.6} \\
Without explicit delegatee responsibility & 50.4{\tiny$\pm$1.8} & 65.0{\tiny$\pm$0.5} \\
Without commitment lifecycle & 56.6{\tiny$\pm$1.4} & 47.8{\tiny$\pm$0.4} \\
Without support dependencies & 60.8{\tiny$\pm$2.1} & 68.9{\tiny$\pm$0.1} \\
Without RCP & 46.7{\tiny$\pm$0.6} & 69.3{\tiny$\pm$0.1} \\
\bottomrule
\end{tabularx}
\caption{Ablation on \textsc{Claw-Eval} with GPT-5.4. Relevant-1 and Random-1 initially assign the task to the most relevant or a randomly selected agent, respectively, instead of all agents.}
\label{tab:illustrative-ablation}
\end{table}

\paragraph{Emergent Coordination Patterns}
To understand the advantages of \method{}, we examine the organizational and coordination patterns that emerge from \method{} trajectories on the Claw-Eval datasets.

Table~\ref{tab:emergent-organization-coordination-main}
(left) demonstrates how \method{} enables
\emph{task-driven self-organization without predefined
workflows or static role assignments}.
Agents autonomously formed diverse coordination
structures, including linear support (35.2\%),
hybrid networks (27.5\%), and specialist hubs
(24.2\%), while adapting delegation to emerging
execution needs (44.0\%).
Expertise roles also emerged dynamically, with agents recruiting missing capabilities (75.8\%) for specialized execution or overlapping capabilities (34.1\%) for verification, coverage, and load sharing. These findings highlight that \emph{\method{} enables adaptive self-organization}, allowing coordination structures, delegation strategies, and expertise roles to evolve under stable responsibility constraints. This reduces reliance on manually engineered
workflows and role assignments as tasks and
agent capabilities evolve.

Table~\ref{tab:emergent-organization-coordination-main}
(right) further shows that adaptive organization
does not necessarily require dense communication.
Agents exhibited \emph{low communication breadth but high coordination depth}, with a mean directed peer-edge
density of 34.25\% and 4.53 exchanges per active
edge.
Crucially, coordination is grounded in
environmental evidence with 94.5\% of peer reports
were environment-grounded, and 84.6\% of
trajectories synthesized evidence across agents.
Moreover, \emph{collective agreement was coupled
to executable task progress rather than
communication alone}: rejected actions prompted
revision (60.4\%), while unresolved dependencies
prevented premature task closure (35.2\%).
Together, these behaviors illustrate how
\method{} supports \emph{accountable decentralized
coordination} through evidence exchange,
action revision, and dependency-aware completion,
without mandatory centralized routing or
global voting.
Under bounded peer degree and exchange depth,
communication can scale as $O(n)$, compared
with $O(dn^2)$ for dense peer interaction with
$d$ events per edge.
These structural bounds suggest a potential
scalability advantage without requiring
predefined communication topologies.

\begin{table}[!t]
\centering
\tiny
\setlength{\tabcolsep}{2pt}
\renewcommand{\arraystretch}{0.9}
\setlength{\aboverulesep}{0.5pt}
\setlength{\belowrulesep}{0.5pt}

\begin{minipage}[t]{0.40\linewidth}
\vspace{0pt}
\begin{tabularx}{\linewidth}{@{}>{\raggedright\arraybackslash}Xr@{}}
\toprule
\textbf{Observed behavior} & \textbf{Rate (\%)} \\
\midrule

\multicolumn{2}{@{}p{\linewidth}@{}}{\textbf{Emergent task architectures}} \\
Linear / sequential support & 35.2 \\
Hybrid network & 27.5 \\
Specialist hub & 24.2 \\
Delegation star & 9.9 \\
Hierarchical cascade & 3.3 \\

\midrule
\multicolumn{2}{@{}p{\linewidth}@{}}{\textbf{Adaptive delegation}} \\
Needs discovered during execution & 44.0 \\
Before execution & 39.6 \\
Both early and during execution & 16.5 \\

\midrule
\multicolumn{2}{@{}p{\linewidth}@{}}{\textbf{Emergent expertise roles}} \\
Complementary capabilities & 75.8 \\
Overlapping capabilities for verification,
coverage, or load sharing & 34.1 \\

\bottomrule
\end{tabularx}
\end{minipage}%
\hfill
\begin{minipage}[t]{0.57\linewidth}
\vspace{0pt}
\begin{tabularx}{\linewidth}{@{}>{\raggedright\arraybackslash}Xr@{}}
\toprule
\textbf{Observed behavior} & \textbf{Rate / Value} \\
\midrule

\multicolumn{2}{@{}p{\linewidth}@{}}{\textbf{Low breadth, high depth}} \\
Directed peer-edge density
& 34.25 (16.67--83.33) \\
Exchanges per active edge
& 4.53 (med. 3.25; max 14) \\

\midrule
\multicolumn{2}{@{}p{\linewidth}@{}}{\textbf{Evidence-grounded collaboration}} \\
Environment-grounded peer reports & 94.5 \\
Multi-agent evidence synthesis & 84.6 \\

\midrule
\multicolumn{2}{@{}p{\linewidth}@{}}{\textbf{Collective agreement}} \\
Rejection triggers action revision & 60.4 \\
Dependencies prevent premature closure & 35.2 \\

\midrule
\multicolumn{2}{@{}p{\linewidth}@{}}{\textbf{Conditional communication complexity}} \\
Bounded peer degree and exchanges & $O(n)$ \\
Dense peer interaction & $O(dn^2)$ \\

\bottomrule
\end{tabularx}
\end{minipage}

\caption{\textbf{Emergent self-organization and local coordination.} Complexity bounds are structural, with $d$ events per edge.}
\label{tab:emergent-organization-coordination-main}
\vspace{-1em}
\end{table}

\section{Related Work}
\label{sec:related}

\paragraph{Multi-agent Coordination} Classical multi-agent coordination formalizes collective activity through joint intentions and responsibilities \citep{levesque1990acting,cohen1991teamwork,jennings1993commitments,jennings1999joint}, social commitments \citep{singh1999commitments,yolum2002flexible,chopra2009alignment}, and protocols governing roles and dependencies \citep{grosz1996collaborative,tambe1997flexible,singh2007formalizing}.  Recent LLM-based MAS operationalize collaboration through designed role-based workflows \citep{hong2024metagpt,wu2023autogen,chen2024agentverse} or learned orchestration structures \citep{zhang2025aflow,zhuge2024gptswarm,zhang2025agentprune}, with dependencies governed by the resulting workflow. Building on both lines of work, \method{} extends social commitments with explicit delegator responsibility and operationalizes joint responsibility through clearly defined obligations%
, maintained across recursive delegation.

\paragraph{LLM-based Decentralized MAS} %
enable scalable and adaptive coordination through local decision-making, with recent work exploring structured communication and collective reasoning \citep{zhang2024machinesom,li2024sparsemad,lee2026robust}, adaptive planning and routing \citep{zhou2025reso,liu2024dylan,yang2025agentnet,yang2026birouter}, and inference-time control \citep{stolfo2025activation}. However, decentralization alone does not explicitly assign responsibility for monitoring dependencies, verifying outcomes, and recovering from failures. \method{} addresses this gap by making each delegator a local orchestrator responsible for its delegated work, collectively enabling decentralized coordination while preserving agent autonomy.

\section{Conclusion}

We introduced a responsibility-centered coordination mechanism for decentralized MAS. By explicitly assigning responsibility for task fulfillment and local orchestration to involved parties, and orienting its decision-making and communication toward fulfilling those responsibilities, \method{} enables autonomous and robust coordination and outperforms strong centralized and decentralized baselines. Our results further show that local orchestration enables agents to detect harmful actions and execution failures, effectively initiate local recovery, and adapt to changes, allowing diverse coordination patterns to emerge without prior specification and improving safety, adaptability, and resilience. By bounding orchestration and communication to local commitments, \method{} reduces coordination overhead and brings communication costs closer to linear $O(n)$ scaling.

\section*{Limitations}

Our evaluation covers multiple reasoning, coding, and executable agent environments, two executor models, and team configurations of different sizes. Nevertheless, these settings cannot represent every task, model, or operating environment in which multi-agent coordination may be deployed. Future work should examine broader deployment settings and more ambitious problems.

\section*{Ethical Considerations}

Our study has several ethical considerations. First, enabling more autonomous multi-agent coordination may reduce direct human oversight and could amplify harmful actions when agents operate with incorrect assumptions, unreliable evidence, or inappropriate permissions. Although \method{} makes operational responsibility explicit and constrains interventions by agent roles, authority, and agreed terms, it does not replace human accountability or guarantee safe behavior. Deployments in high-stakes settings should therefore retain appropriate access controls, auditing, and human oversight.

Second, evaluating multiple models, coordination methods, and repeated runs requires substantial computation, with associated energy consumption and carbon emissions. We limit this cost by using fixed evaluation sets, reusing execution traces where appropriate, and reporting results over three runs rather than conducting unrestricted large-scale searches.

\bibliography{custom}
\clearpage
\appendix
\section{Additional Evaluation Details}
\label{app:evaluation-details}

\subsection{Commitment Lifecycle Statuses}
\label{app:commitment-record}

\begin{figure}[t]
    \centering
    \includegraphics[
        width=\linewidth
    ]{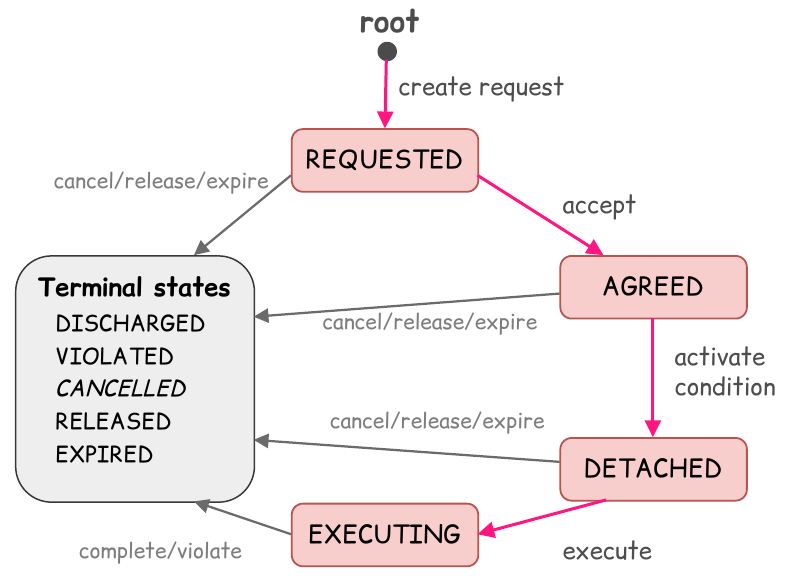}
    \caption{Commitment lifecycle and state transitions relevant to responsibility in \method{}.}
    \label{fig:commitment-lifecycle}
\end{figure}

Table~\ref{tab:commitment-lifecycle} defines the nine commitment lifecycle statuses, and Figure~\ref{fig:commitment-lifecycle} summarizes their transitions. A request records a proposal; acceptance of the same proposal version establishes a binding commitment. The nominal exchange between the parties is shown in Figure~\ref{fig:commitment-handshake} in the main text.

\begin{table*}[t]
\centering
\small
\begin{tabular}{p{1.7cm} p{10.8cm}}
\toprule
\textbf{Status} & \textbf{Description} \\
\midrule

\textit{REQUESTED} &
The delegator records the proposed goal and terms and sends the request. No obligation exists yet. \\

\textit{AGREED} &
The delegatee accepts the proposal. The commitment is now binding, although its activation conditions may not yet hold. \\

\textit{DETACHED} &
The activation conditions in $\tau$ hold, making the obligation to achieve $g$ active. \\

\textit{EXECUTING} &
Records that the delegatee has started work and reported progress. \\

\textit{DISCHARGED} &
The delegator has verified that the outcome satisfies the agreed completion criteria. \\

\textit{VIOLATED} &
The delegator has verified that a due outcome was not achieved or that an agreed constraint was breached. \\

\textit{CANCELLED} &
The delegator has terminated the commitment before verified fulfilment. \\

\textit{RELEASED} &
The delegator has released the delegatee from the obligation without recording a violation. \\

\textit{EXPIRED} &
The deadline or validity window has ended before the commitment was resolved. \\

\bottomrule
\end{tabular}
\caption{Commitment lifecycle states and their operational semantics in \method{}.}
\label{tab:commitment-lifecycle}
\end{table*}

\paragraph{Status transitions.}
An event $e_t$ identifies its actor, agreement version, event type, and supporting evidence. The transition uses the current state, the event, and the accepted terms $\tau$:
\[
s_{t+1}(C)=\delta_C\bigl(s_t(C),e_t;\tau\bigr).
\]
A valid acceptance establishes \textsc{Agreed}; evidence that $\alpha$ holds permits \textsc{Detached}; and a report that execution has begun permits \textsc{Executing}. Acceptance and activation may occur together when $\alpha$ already holds. A completion submission provides evidence for review without introducing another lifecycle state. It permits \textsc{Discharged} only when the delegator verifies that its evidence satisfies $\beta$. Further work may remain necessary when a submission is insufficient; insufficient evidence alone is not a verified violation.

A verified unmet due outcome or breached constraint permits \textsc{Violated}. Authorized termination before verified fulfilment permits \textsc{Cancelled}; release without recording a violation permits \textsc{Released}. An unresolved record reaches \textsc{Expired} when its agreed deadline or validity window ends. Each transition checks the applicable authority, version, conditions, and evidence before taking effect. Invalid or unauthorized events do not alter the accepted record. Terminal records remain available for assessing effects on other commitments.

\subsection{RCP Processing and Transport}
\label{app:rcp-details}

\paragraph{Processing rules.}
An agent processes messages against its local coordination record, checking identity, version, sender authority, and any required evidence. Questions, clarifications, progress reports, exceptions, and outcome evidence inform the local decision policy \citep{singh2007formalizing}. Under local orchestration, the delegator updates the lifecycle state on the commitment edge only when validated events, conditions, and evidence satisfy the transition guards in Appendix~\ref{app:commitment-record}. Questions and reports do not by themselves change its lifecycle state.

Duplicate message identifiers are idempotent. A stale version returns an error identifying the current version. Ordering is required within the affected agreement, without imposing a total order across unrelated commitments. These rules support local processing; they do not require agents to share a globally synchronized execution trace.

\paragraph{Transport binding.}
RCP separates message semantics from transport, following the interoperability approach of MCP and A2A \citep{hou2025mcp,a2aproject2026}. Its envelopes can be carried over an authenticated bidirectional channel through, for example, a JSON-RPC binding, an A2A extension, or a local message bus. A binding may change serialization and delivery while preserving commitment identity, versioning, and correlation. Each agent retains its own planner, tools, and internal execution state.

\subsection{Benchmark Details}
\label{app:benchmark-details}
\paragraph{AIME25.}
\textsc{AIME25} is a competition-mathematics benchmark designed to test multi-step mathematical reasoning with objectively verifiable answers \citep{aime25}. Because it requires no tool interaction or changing environment state, it serves as a capability control: improvements should reflect useful coordination, while regressions reveal coordination overhead that disrupts an executor's underlying reasoning.

\paragraph{HumanEval.}
\textsc{HumanEval} contains 164 hand-written Python synthesis problems specified by function signatures and docstrings and evaluated by unit tests \citep{chen2021humaneval}. Pass@1 therefore measures functional correctness rather than textual similarity.

\paragraph{Claw-Eval.}
\textsc{Claw-Eval} comprises 300 human-verified tasks across general service orchestration, multimodal interaction, and multi-turn professional dialogue, with 2,159 fine-grained rubric items \citep{ye2026claweval}. It records execution traces, audit logs, and environment snapshots and separately evaluates completion, safety, and robustness. This trajectory-grounded evidence is essential for our study because it reveals not only whether the final task succeeds, but whether agents detect consequential failures, communicate them, monitor the affected work, and intervene safely under controlled execution failures.

\paragraph{Scoring.}
For task $t$, let $r_{t\ell}\in[0,1]$ be the result of rubric item $\ell$ with weight $w_{t\ell}$. Completion is the normalized weighted rubric score, $s_{\mathrm{comp}}(t)=\sum_{\ell}w_{t\ell}r_{t\ell}/\sum_{\ell}w_{t\ell}$. The official score for one attempt is
\[
s_t=s_{\mathrm{safe}}(t)\left(0.8\,s_{\mathrm{comp}}(t)+0.2\,s_{\mathrm{rob}}(t)\right),
\]
where the safety score is a multiplicative gate. If $\mathcal{T}_{\mathrm{err}}$ is the set of tool types that encounter an injected error and $\mathcal{T}_{\mathrm{rec}}$ is the subset later returning a successful response, then $s_{\mathrm{rob}}=|\mathcal{T}_{\mathrm{rec}}|/|\mathcal{T}_{\mathrm{err}}|$, or $1$ when no error is injected. For each run $r$, the dataset score is the mean over all tasks,\[\mathrm{Score}^{(r)}=\frac{100}{|\mathcal{T}|}\sum_{t\in\mathcal{T}}s_t^{(r)}.\] We report the mean $\pm$ standard deviation of $\mathrm{Score}^{(r)}$ over 3 different runs.

\paragraph{Claw-Eval-Live.}
\textsc{Claw-Eval-Live} evaluates 105 end-to-end tasks spanning controlled business services and local workspace repair \citep{li2026clawevallive}. Its task mixture is derived from refreshable public workflow-demand signals, while each release remains reproducible through a time-stamped snapshot with fixed fixtures, services, workspaces, and graders. Execution traces, audit logs, service state, and post-run artifacts verify that work was actually performed. The benchmark therefore tests whether coordination generalizes beyond a fixed task set and remains effective as workflow demand and operating conditions evolve.

\paragraph{Scoring.}
Each task-specific grader produces a completion score $s_{t,m}\in[0,1]$ from execution evidence, using deterministic checks when possible and structured LLM judgment only for semantic rubric items. For each run $r$, the dataset score is the mean over all tasks,
\[
\mathrm{Score}^{(r)}=\frac{100}{|\mathcal{T}|}\sum_{t\in\mathcal{T}}s_t^{(r)}.
\]
We report the mean $\pm$ standard deviation of $\mathrm{Score}^{(r)}$ over 3 different runs.

Together, the four benchmarks separate underlying reasoning and coding capability from executable coordination and adaptation. AIME25 and HumanEval expose unnecessary coordination overhead, Claw-Eval measures safe and robust behavior under observable failures, and Claw-Eval-Live tests transfer to evolving workflows.

\begin{table}[h]
\centering
\small
\resizebox{\columnwidth}{!}{
\begin{tabular}{lll}
\toprule
\textbf{Dataset} & \textbf{Setting} & \textbf{Primary metric} \\
\midrule
AIME25 & Mathematical reasoning & Accuracy \\
HumanEval & Code generation & Pass@1 \\
Claw-Eval & Executable agent tasks & Score, completion, safety, responsibility \\
Claw-Eval-Live & Dynamic executable tasks & Score \\
\bottomrule
\end{tabular}}
\caption{Evaluation benchmarks, task settings, and primary metrics.}
\label{tab:benchmark-details}
\end{table}

\subsection{Failure Signals}
\label{app:failure-signals}
We group them as follows:
\begin{itemize}
    \item \textbf{Execution:} HTTP status $\geq 400$, non-zero shell exit, timeout, \texttt{rejected}, or \texttt{denied}.
    \item \textbf{Resource/state:} \texttt{out\_of\_stock}, \texttt{unavailable}, \texttt{missing}, \texttt{not\_found}, or \texttt{blocked}.
    \item \textbf{Operational:} \texttt{overdue}, \texttt{expired}, \texttt{cancelled}, \texttt{failed}, \texttt{critical}, \texttt{degraded}, or \texttt{unhealthy}.
    \item \textbf{Validation/safety:} \texttt{invalid}, \texttt{unsafe}, \texttt{violation}, or false \texttt{success}, \texttt{health}, or \texttt{valid} flags.
\end{itemize}
Signals are deduplicated within each task/trial by $(\texttt{tool},\texttt{entity},\texttt{field},\texttt{value})$. Each method uses its own observed-failure denominator rather than a cross-method union.

\subsection{Baselines}
\label{app:baseline-details}

We consider three families of multi-agent baselines based on their primary coordination mechanisms.
\emph{Decentralized communication} includes \textsc{MAS-Broadcast}, which enables all-to-all information sharing; \textsc{MachineSoM}, which supports peer collaboration through debate and reflection; \textsc{Sparse MAD}, which employs sparse peer-to-peer communication; and \textsc{AgentNet}, which enables decentralized task routing over a dynamically evolving interaction graph \citep{zhang2024machinesom,li2024sparsemad,yang2025agentnet}.
\emph{Hand-designed coordination} includes \textsc{MetaGPT}, \textsc{AutoGen}, and \textsc{AgentVerse}, which organize agent interactions through predefined roles, workflows, or conversation protocols \citep{hong2024metagpt,wu2023autogen,chen2024agentverse}.
\emph{Automatically optimized coordination} includes \textsc{AFlow}, \textsc{GPTSwarm}, and \textsc{AgentPrune}, which optimize agentic workflows or communication structures through search, graph optimization, or pruning \citep{zhang2025aflow,zhuge2024gptswarm,zhang2025agentprune}.
On Who\&When, we include trained attribution models from Aegis \citep{kong2026aegis} and training-free LLMs. On ATBench-C, we include guard models \citep{meta2024llamaguard3,meta2025llamaguard4,zhao2025qwen3guard,chen2025shieldagent,liu2026agentdog} and general-purpose LLMs.

\subsection{Implementation Details}
\label{app:implementation-details}

All main experiments use sixteen-agent teams with either \textsc{Qwen3-4B-Instruct-2507} \citep{yang2025qwen3} or \textsc{GPT-5.4} \citep{openai2026gpt54}. In agentic environments, each agent receives only its assigned subset of tools and permissions, while the complete capability set is available only to the team; completing a task therefore requires agents to communicate and delegate work to one another. For \method{}, each agent can discover and directly contact any other team member through a directory of agent identities, capabilities, and endpoints. We set the per-round candidate limit to $K=2$ and the discovery round budget to $L=1$ per attempt. Each attempt therefore contacts at most two candidates in one round and stops on agreement, candidate exhaustion, or the round budget. Recontact is allowed, with at least one untried candidate included whenever available; subsequent attempts are triggered by revised requirements or changed conditions. We use temperature $0.7$, top-$p$ $0.8$, at most 220 agent events, and maximum output lengths of 32,768 tokens; Qwen additionally uses top-$k$ $20$. Each configuration is evaluated over 3 different runs, and we report the mean and standard deviation. For Qwen, each agent uses a separate base-model instance hosted on a dedicated NVIDIA A100 80GB GPU and accessed via an API; GPT-5.4 uses its provider API. Latency is measured as end-to-end wall-clock time per task, including cross-server API overhead. It is therefore used for relative comparisons among methods under the same serving setup, with Qwen and GPT-5.4 reported separately.

\end{document}